# On the Performance under Hard and Soft Bitwise Mismatched-Decoding

**Tsuyoshi Yoshida[1,2], Mikael Mazur[3], Jochen Schröder[3], Magnus Karlsson[3], and Erik Agrell[3]**

[1] *Information Technology R&D Center, Mitsubishi Electric Corporation, Kamakura, 247-8501 Japan*
[2] *Graduate School of Engineering, Osaka University, Suita, 505-0871 Japan*
[3] *Fiber-Optic Communications Research Center (FORCE), Chalmers University of Technology, SE-41296, Gothenburg, Sweden*
*Yoshida.Tsuyoshi@ah.MitsubishiElectric.co.jp, yoshida.tsuyoshi@opt.comm.eng.osaka-u.ac.jp*

**Abstract:** We investigated a suitable auxiliary channel setting and the gap between Q-factors with hard and soft demapping. The system margin definition should be reconsidered for systems employing complex coded modulation with soft forward error correction. © 2020 The Author(s)
**OCIS codes:** (060.2330) Fiber Optics Communication; (060.4080) Modulation

## 1. Introduction

Spectrally-efficient high capacity communication is required in metro-to-long-haul optical fiber communications. Key technologies in those systems are quadrature amplitude modulation (QAM), forward error correction (FEC), and probabilistic shaping (PS) [1,2]. Deployable systems tend to employ binary FEC, and then the bits or bitwise L-values (log-likelihood ratios) are recovered in the demapping from the received symbols. The processing and the design of the demapping is more complex with multilevel modulation such as higher-order QAM or formats using PS than the conventional modulation formats such as quaternary phase-shift keying. How to select the channel assumed in the demapping process, the so called *auxiliary channel*, is then a nontrivial task.

With a bitwise receiver, the performance has been quantified by bit error rate (BER), generalized mutual information (GMI) [3,4], or asymmetric information (ASI) [5]. In most works with bit-interleaved coded modulation (BICM) [6] or BICM with PS [1,2], the auxiliary channel would be assumed to be matched to the channel at least approximately when deriving the GMI. Recently, works on multilevel (binary) coding (MLC) with a bitwise receiver have seen increased attention, since such schemes may show an appealing balance of performance versus power consumption [7–10]. They typically protect the least significant bit tributary by a soft FEC and the others by hard FEC with the help of multistage decoding. The maximum achievable rate of MLC is the mutual information (MI) between transmitted and received symbols [11]. In any coded modulation scheme, deployed systems are operated with a reasonable margin above the FEC threshold. It is essential to measure the margin from live traffic without knowledge of the transmitted bits. The pre-FEC BER can be estimated from the fraction of flipped bits in the FEC decoding, converted to Q-factor, and used in the system margin description. Alternatively, the ASI can be estimated blindly [12].

Here we have two issues. The first issue is that the matched decoding assumption is questionable to quantify the performance in deployable systems in the presence of a system margin. The second issue is that the prediction of post-soft FEC BER in MLC has not been sufficiently studied. Thus, in this paper, we report how the Q-factors converted from pre-FEC BER/ASI based on hard/soft bitwise demapping are influenced by mismatched decoding via simulation and experiment. We study the validity of the matched decoding assumption for deployable systems and the dependence on different coded modulation schemes such as BICM and MLC. Note that this work focuses on the true performance behavior under various mismatched decoding. We will not use any blind performance monitoring methods like [12] to avoid the uncertainty due to the estimation error.

## 2. System model and performance metrics

Fig. 1 shows the system model and relevant performance metrics in this work. Source bits $S$ are encoded into the bits $A$ by a PS encoder. The PS-encoded bits $A$ are further encoded systematically to generate bits $B$. The FEC-encoded bits $B$ are mapped to the symbols $X$. The channel is simulated as a discrete memoryless channel, analytically derived by quantizing the output $Y$ of a Gaussian channel with a given the signal-to-noise ratio (SNR), $SNR_{tr}$. The received symbols $Y$ are demapped with the auxiliary channel $q_{Y|B}$, which is assumed to be a quantized Gaussian channel with the SNR $SNR_{aux}$. Then bitwise $a$

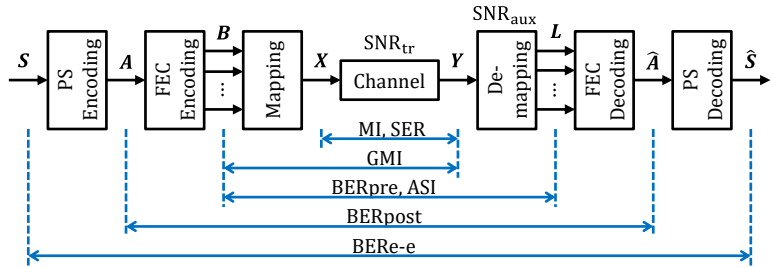

Fig. 1. A system model and relevant performance metrics in coded modulation systems.

*posteriori* L-values $\boldsymbol{L}$ are derived and the L-values are used for the FEC decoding to the estimated payload $\widehat{\boldsymbol{A}}$. Finally PS decoding is performed to recover the estimated source bits $\widehat{\boldsymbol{S}}$. The MI between $\boldsymbol{X}$ and $\boldsymbol{Y}$ gives an achievable rate with MLC, the symbol error rate (SER) describes the channel performance simply, and the end-to-end BER after PS decoding ($\text{BER}_{e-e}$) is critical to the system design. In this work, to analyze the influence of the demapping, we focus on the pre-FEC BER ($\text{BER}_{pre}$), the ASI [5], and the post-FEC BER ($\text{BER}_{post}$). The ASI is defined as $1 - h(L_a) - h(|L_a|)$, where $L_a = (-1)^b L$, $b$, and $h(\cdot)$ are the symmetrized L-value, the transmitted bit, and differential entropy, resp. We consider the Q-factors $Q_{\text{BER}} = \sqrt{2}\text{erfc}^{-1}(2 \cdot \text{BER}_{pre})$ and $Q_{\text{ASI}} = J^{-1}(\text{ASI})$ with the hard and soft demapping, where $J^{-1}(\cdot)$ is the inverse J-function [13]. We need to set an SNR for the demapping, $\text{SNR}_{aux}$, and the performance is maximized when $\text{SNR}_{aux} = \text{SNR}_{tr}$. The ASI under matched decoding is equivalent to the normalized GMI as shown in [5,12]. We will investigate how the Q-factors depend on $\text{SNR}_{tr}$ and $\text{SNR}_{aux}$.

## 3. Performance under various auxiliary channels

We simulated the performance over the Gaussian channel in the following two cases: (a) $\text{SNR}_{aux} = \text{SNR}_{tr}$ and (b) $\text{SNR}_{aux} = \text{SNR}_{lim}$, where $\text{SNR}_{lim}$ denotes the SNR limit at the FEC threshold for a given communication scheme. We consider an inner soft FEC and an outer hard FEC. The received symbol and the L-values were quantized to 7 bits and 5 bits, resp. We employ the DVB-S2 low-density parity-check FEC code having a soft FEC code rate $R_c$ of 4/5 and 20 decoding iterations. $\text{SNR}_{lim}$ is assumed to be the SNR at the soft FEC decoding output BER $\approx 10^{-4}$.

In Fig. 2 we examine BICM PS-64-QAM having the same information rate as 32-QAM. Under matched decoding in Fig. 2(a), $Q_{\text{ASI}}$ is almost equal to $Q_{\text{BER}}$ at high SNR, or slightly larger at low SNR. Under mismatched decoding, by assuming $\text{SNR}_{aux}$ to be fixed at the FEC threshold, in Fig. 2(b), the $Q_{\text{ASI}}$ degradation is larger when there is a large gap between $\text{SNR}_{aux}$ and $\text{SNR}_{tr}$ but there is almost no degradation in $Q_{\text{BER}}$. When estimating the performance from the bit flipping count in the FEC decoder, the gap between $Q_{\text{BER}}$ and $Q_{\text{ASI}}$ must be accounted for.

Here we consider the case when $\text{SNR}_{tr}$ is suddenly reduced to $\text{SNR}_{lim}$ due to some external factor. In the auxiliary channel adaptation case (a), $\text{SNR}_{aux}$ cannot immediately follow but remains at the previous value, which is now mismatched. This causes a significant degradation of $Q_{\text{ASI}}$ and post-soft FEC BER as shown with purple and blue arrows in Fig. 2(a). The outer hard FEC cannot correct all errors, and $\text{SNR}_{lim}$ cannot be used as an FEC threshold. In the fixed auxiliary channel case (b), $\text{SNR}_{aux}$ becomes matched to $\text{SNR}_{tr}$, when the SNR is suddenly reduced, and we get the expected $Q_{\text{ASI}}$ and post-soft FEC BER at $\text{SNR}_{tr} = \text{SNR}_{lim}$. Thus to fix the auxiliary channel at the FEC threshold is a safe and reasonable option to avoid the aforementioned SNR limit degradation. Then how to quantify the system margin will be another issue. While $\Delta Q_{\text{BER}} = Q_{\text{BER}} - Q_{\text{BER,th}}$ is easily estimated from the bit flipping count in the FEC decoder, $\Delta Q_{\text{ASI}} = Q_{\text{ASI}} - Q_{\text{ASI,th}}$ and $\Delta\text{SNR} = \text{SNR}_{tr} - \text{SNR}_{lim}$ are more reliable but their estimation complexity and accuracy would be issues, since $Q_{\text{BER,th}}$ and $Q_{\text{ASI,th}}$ are Q-factors at $\text{SNR}_{tr} = \text{SNR}_{lim}$.

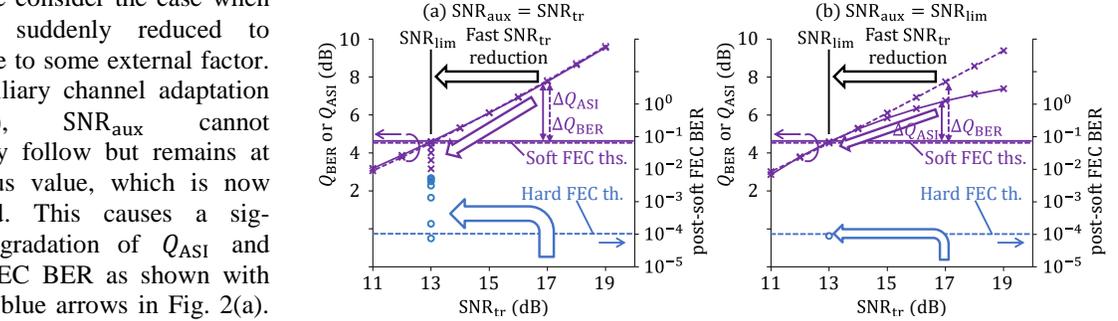

Fig. 2. Simulated performance of BICM PS-64-QAM in cases of (a) $\text{SNR}_{aux} = \text{SNR}_{tr}$ and (b) $\text{SNR}_{aux} = \text{SNR}_{lim}$ (13.0 dB here). The solid and dotted lines show $Q_{\text{ASI}}$ and $Q_{\text{BER}}$, resp.

To see how $Q_{\text{BER}}/Q_{\text{ASI}}$ work as FEC thresholds under mismatched decoding, we simulated post-soft FEC BER for various signals as a function of (a) relative SNR, (b) $Q_{\text{BER}}$, and (c) $Q_{\text{ASI}}$ around $\text{SNR}_{lim}$, as shown in Fig. 3.

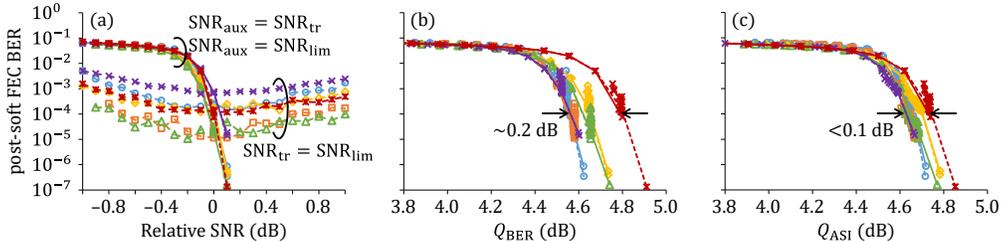

Fig. 3. Simulated post-soft FEC BER as a function of (a) relative SNR, (b) $Q_{\text{BER}}$, and (c) $Q_{\text{ASI}}$ for three demapping cases. Solid line: $\text{SNR}_{aux} = \text{SNR}_{tr}$, dotted line: $\text{SNR}_{aux} = \text{SNR}_{lim}$, and dashed line: $\text{SNR}_{tr} = \text{SNR}_{lim}$. Blue: 128-QAM, orange: 64-QAM, yellow: 32-QAM, green: 16-QAM, purple: BICM PS-64-QAM, red: MLC PS-64-QAM.

$\text{SNR}_{\text{lim}}$ is 10.9 dB, 13.9 dB, 16.3 dB, 19.0 dB, 13.0 dB, and 13.4 dB for 16, 32, 64, 128-QAM, BICM PS-64-QAM, and MLC PS-64-QAM, resp. In this MLC, soft FEC protects the least significant bits only, the parity bits are placed only on the sign bit, and multistage decoding is performed [10]. Under $\text{SNR}_{\text{aux}} = \text{SNR}_{\text{tr}}$ (solid line) and $\text{SNR}_{\text{aux}} = \text{SNR}_{\text{lim}}$ (dotted line), the relative SNR is defined as $\text{SNR}_{\text{tr}}/\text{SNR}_{\text{lim}}$. Under $\text{SNR}_{\text{tr}} = \text{SNR}_{\text{lim}}$, the relative SNR is $\text{SNR}_{\text{aux}}/\text{SNR}_{\text{lim}}$. In Fig. 3(a), there is no significant difference between the two demapping cases of $\text{SNR}_{\text{aux}} = \text{SNR}_{\text{tr}}$ and $\text{SNR}_{\text{aux}} = \text{SNR}_{\text{lim}}$, while post-soft FEC BER is worse with the demapping of $\text{SNR}_{\text{tr}} = \text{SNR}_{\text{lim}}$ (dashed line) and the larger gap between $\text{SNR}_{\text{tr}}$ and $\text{SNR}_{\text{aux}}$. From Figs. 3(b) and 3(c), $Q_{\text{ASI}}$ is better correlated to post-soft FEC BER than $Q_{\text{BER}}$ is, because $Q_{\text{ASI}}$ quantifies soft information quality. For example, at a post-soft FEC BER of $10^{-4}$, the Q-factor differences are 0.2 dB and <0.1 dB for $Q_{\text{BER}}$ and $Q_{\text{ASI}}$, resp. Especially the difference between BICM and MLC is relevant, which comes from the non-Gray labeling used for set partitioning in MLC.

We also analyzed experimental data captured in the setup shown in [11] under matched ($\text{SNR}_{\text{aux}} = \text{SNR}_{\text{tr}}$) or mismatched ($\text{SNR}_{\text{aux}} = \text{SNR}_{\text{lim}}$) decoding. The 24 Gbaud dual-polarized signal was generated by an arbitrary waveform generator. The symbol entropy and the information rate per QAM symbol were 5.2 bit and 4.0 bit for BICM PS-64-QAM and 4.6 bit and 4.1 bit for MLC PS-64-QAM, resp. The 51 tones were multiplexed at 25 GHz spacing. A recirculating loop setup with two 80 km standard single-mode fiber spans was used to vary the transmission distance. The received signal was coherently detected, sampled, and the symbols were recovered by pilot-aided offline processing [14]. The soft FEC decoding was performed with the method in [15]. Fig. 4 shows the Q-factors and post-soft FEC BER for PS-64-QAM with (a) BICM and (b) MLC. $Q_{\text{ASI}}$ with mismatched decoding is smaller than the other Q-factors except at $Q_{\text{ASI}} \approx 4.8$ dB. The post-soft FEC BER is below the assumed threshold BER ($10^{-4}$) input to hard FEC decoding in the regime of $Q_{\text{ASI}} > 4.8$ dB.

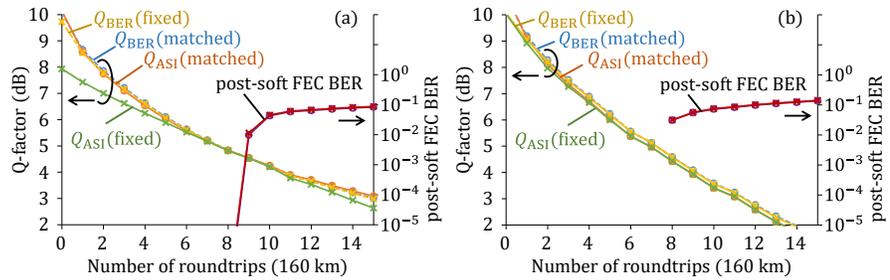

Fig. 4. Experimental Q-factors and post-soft FEC BER with matched (circle) or fixed (cross) auxiliary channel for PS-64-QAM with (a) BICM and (b) MLC. There was no residual error after FEC decoding over 1000 DVB-S2 FEC codewords at 7 or fewer roundtrips in both the BICM and MLC cases.

## 5. Conclusions

We studied how mismatched decoding reduces the soft Q-factor and leads to a gap between Q-factors from hard/soft bitwise demapping, through simulations and experiments. A safe demapping scheme in deployable (=bitwise decoding with limited complexity) systems is to fix the SNR of the auxiliary channel to that of a given FEC threshold, which is independent of the channel SNR. The relationship between hard and soft Q-factors also depends on the QAM order and coded modulation scheme such as BICM or MLC. Even then, the soft Q-factor works better than the hard one, not only in BICM/BICM-PS but also for MLC, to predict the BER after soft FEC decoding even under significantly SNR-mismatched decoding.

**Acknowledgments**
This work was partly supported by "Massively Parallel and Sliced Optical Network," the Commissioned Research of National Institute of Information and Communications Technology (NICT), Japan. We thank researchers in Mitsubishi Electric Research Laboratories for discussions about the Q-factor definition. We also thank Assoc. Prof. Koji Igarashi and Prof. Kyo Inoue of Osaka University for assistance in the research.